\documentclass[oneside,article]{memoir}

\usepackage{amsmath}         % align-like environments
\usepackage{amsthm,thmtools,thm-restate} % theorem-like environments
\usepackage{enumitem}        % enumerate-like environments

\usepackage{rotating}

\usepackage[final]{listings}
\usepackage{bbm}
\usepackage{xcolor}

\definecolor{codegreen}{rgb}{0,0.6,0}
\definecolor{codegray}{rgb}{0.5,0.5,0.5}
\definecolor{codepurple}{rgb}{0.58,0,0.82}
\definecolor{backcolour}{rgb}{0.95,0.95,0.92}

\lstdefinestyle{mystyle}{
    backgroundcolor=\color{backcolour},   
    commentstyle=\color{codegreen},
    keywordstyle=\color{magenta},
    numberstyle=\tiny\color{codegray},
    stringstyle=\color{codepurple},
    basicstyle=\ttfamily\footnotesize,
    breakatwhitespace=false,         
    breaklines=true,                 
    captionpos=b,                    
    keepspaces=true,                 
    numbers=left,                    
    numbersep=5pt,                  
    showspaces=false,                
    showstringspaces=false,
    showtabs=false,                  
    tabsize=2
}

\usepackage{algorithm}
\usepackage{algpseudocode}
\usepackage{algorithmicx}
\usepackage{threeparttable}
\usepackage{booktabs}
\usepackage{float}

\makeatletter
\providecommand{\ALG@name}{Algorithm}   % guard: normally set by algorithm.sty
\newenvironment{balgorithm}
  {\par\addvspace{\topsep}%
   \refstepcounter{algorithm}%
   \renewcommand{\caption}[1]{%
     \noindent{\bfseries \ALG@name~\thealgorithm.\ }##1\par\nobreak\vskip 3pt}%
  }
  {\par\addvspace{\topsep}}
\makeatother

\providecommand{\needspace}[1]{}

\usepackage{tikz}
\usetikzlibrary{matrix,arrows,positioning}

\usepackage{pgfplots}
\pgfplotsset{compat=1.17}
\usepgfplotslibrary{groupplots}

\definecolor{dhcol}{RGB}{45,110,165}   % discrete / RedZeD_DH / RedZeD_PDH
\definecolor{vrcol}{RGB}{215,140,40}   % simplicial / RedZeD_VR
\definecolor{qcol}{RGB}{170,60,50}     % Quot_DH
\definecolor{acol}{RGB}{40,130,90}     % accent

\pgfplotsset{
  redzedplot/.style={
    width=0.47\textwidth, height=5.2cm,
    ymajorgrids, grid style={gray!25},
    label style={font=\small}, tick label style={font=\footnotesize},
    title style={font=\small}, legend style={font=\footnotesize, draw=none,
      fill=white, fill opacity=0.85, text opacity=1, cells={anchor=west}},
    every axis plot/.append style={thick},
    log basis y=10,
  },
}

\tikzset{
  dhline/.style={dhcol, mark=*, mark size=1.6pt},
  vrline/.style={vrcol, mark=square*, mark size=1.6pt},
  qline/.style={qcol, mark=triangle*, mark size=2.2pt},
  estline/.style={dashed, mark options={fill=white}},
}
\usepackage{amsfonts}
\usepackage{parskip}
\usepackage{mathtools}
\usepackage{wrapfig}

\usepackage{tikz-cd}

\usepackage{caption, subcaption} % for subfigures

\usepackage[hidelinks]{hyperref}                   % hyperlinks
\usepackage[capitalize,nosort]{cleveref}           % named references

\usepackage{amssymb}
\usepackage[mathscr]{euscript}
\usepackage{relsize}
\usepackage{stmaryrd}
\usepackage{stackengine}

\usepackage{indentfirst} % indent the first paragraph
\usepackage{multicol}

\usepackage{graphicx}
\usepackage{ragged2e} % for nicer left alignment inside columns
\usepackage[inline,nomargin,draft]{fixme}

\isopage[12]

\setlrmargins{*}{*}{1}
\checkandfixthelayout

\counterwithout{section}{chapter}
\usepackage{appendix}

\makeatletter
\providecommand{\abx@aux@refcontext}[1]{}
\providecommand{\abx@aux@cite}[2]{}
\providecommand{\abx@aux@segm}[3]{}
\providecommand{\abx@aux@page}[2]{}
\providecommand{\abx@aux@fnpage}[2]{}
\providecommand{\abx@aux@backref}[5]{}
\providecommand{\abx@aux@defaultrefcontext}[3]{}
\providecommand{\abx@aux@read@bbl@mdfivesum}[1]{}
\providecommand{\abx@aux@read@bblrerun}{}
\providecommand{\abx@aux@sortscheme}[1]{}
\providecommand{\abx@aux@refsection}[1]{}
\providecommand{\abx@aux@number}[2]{}
\makeatother

\makeatletter
\let\dgm\@undefined
\makeatother

  \newcommand{\iso}{\mathrel{\cong}}

  \declaretheorem[style=definition,within=section]{definition}
  \declaretheorem[style=definition,numberlike=definition]{example}

  \declaretheorem[style=plain,numberlike=definition]{theorem}
  \declaretheorem[style=plain,numberlike=definition]{conjecture}

  \declaretheorem[style=plain,numbered=no,name=Theorem]{theorem*}

  \Crefname{corollary}{Corollary}{Corollaries}
  \Crefname{definition}{Definition}{Definitions}
  \Crefname{lemma}{Lemma}{Lemmas}
  \Crefname{proposition}{Proposition}{Propositions}
  \Crefname{remark}{Remark}{Remarks}
  \Crefname{theorem}{Theorem}{Theorems}
  \Crefname{notation}{Notation}{Notations}
  \Crefname{conjecture}{Conjecture}{Conjectures}

  \newlist{axioms}{enumerate}{1}
  \Crefname{axiomsi}{}{}

  \newenvironment{tikzeq*}
  {
    \begingroup
    \begin{equation*}
    \begin{tikzpicture}[baseline=(current bounding box.center)]
  }
  {
    \end{tikzpicture}
    \end{equation*}
    \endgroup
    \ignorespacesafterend
  }

  \tikzset
  {
    diagram/.style=
    {
      matrix of math nodes,
      column sep={4.3em,between origins},
      row sep={4em,between origins},
      text height=1.5ex,
      text depth=.25ex
    },
    over/.style={preaction={draw=white,-,line width=6pt}},
    every to/.style={font=\footnotesize},
    inj/.style={right hook->},
    surj/.style={-{Latex[open]}},
    cof/.style={>->},
    fib/.style={->>},
  }

  \DeclareFontFamily{U}{mathx}{\hyphenchar\font45}

  \DeclareFontShape{U}{mathx}{m}{n}{
    <5> <6> <7> <8> <9> <10>
    <10.95> <12> <14.4> <17.28> <20.74> <24.88>
    mathx10}{}

  \DeclareSymbolFont{mathx}{U}{mathx}{m}{n}

  \DeclareFontFamily{U}{mathb}{\hyphenchar\font45}

  \DeclareFontShape{U}{mathb}{m}{n}{
    <5> <6> <7> <8> <9> <10>
    <10.95> <12> <14.4> <17.28> <20.74> <24.88>
    mathb10}{}

  \DeclareSymbolFont{mathb}{U}{mathb}{m}{n}

  \DeclareMathAccent{\widebar}{0}{mathx}{"73}

  \DeclareMathSymbol{\Rsh}{\mathrel}{mathb}{"E9}

  \DeclareFontFamily{U}{MnSymbolA}{}

  \DeclareFontShape{U}{MnSymbolA}{m}{n}{
    <-6> MnSymbolA5
    <6-7> MnSymbolA6
    <7-8> MnSymbolA7
    <8-9> MnSymbolA8
    <9-10> MnSymbolA9
    <10-12> MnSymbolA10
    <12-> MnSymbolA12}{}

  \DeclareSymbolFont{MnSyA}{U}{MnSymbolA}{m}{n}

  \DeclareMathSymbol{\twoheaddownarrow}{\mathrel}{MnSyA}{27}

  \newcommand{\MSC}[1]{%
    \let\thempfn\relax
    \footnotetext[0]{2020 Mathematics Subject Classification: #1.}
  }

  \newcommand{\dgm}{\textsf{dgm}}

  \newcommand{\keys}{\operatorname{keys}}

  \renewcommand{\dgm}{\operatorname{dgm}}

  \newcommand{\bigtriangle}{\scalebox{1.5}{$\triangle$}}

\tikzstyle{vertex}=[circle, draw, minimum size=7pt, inner sep=0pt]

\newcommand{\redzed}{{\textsc{RedZeD}}}
\newcommand{\redzedvr}{{\textsc{RedZeD{\textunderscore}VR}}}
\newcommand{\redzeddh}{{\textsc{RedZeD{\textunderscore}DH}}}
\newcommand{\redzedpdh}{{\textsc{RedZeD{\textunderscore}PDH}}}
\newcommand{\bbF}[1]{\mathbb{F}_{#1}}

\newcommand{\image}{\operatorname{im}} % image of a function
\DeclareFontFamily{U}{dmjhira}{}
\DeclareFontShape{U}{dmjhira}{m}{n}{ <-> dmjhira }{}

\author{Sterling Ebel \and Krzysztof Kapulkin \and Nathan Kershaw}

\title{Discrete homology computations by reduction to zero differentials}

\date{\today}

\begin{document}

  \maketitle

\begin{abstract}
   We develop a new algorithm for computing (persistent) discrete homology of graphs using reduction to zero differentials and active enumeration.
   This allows us to compute the fourth homology group of the Greene sphere, along with several previously unknown groups.
   We also show that persistent discrete homology computes faster than simplicial homology of Vietoris--Rips complex in the high-noise non-metric settings, making it a better choice for noisy data sets.
\end{abstract}

%  \setlist[enumerate]{label=(\arabic*)}

% Add content here

\section*{Introduction}

Discrete homology is a homology theory for graphs.
Its chain module in degree $n$ is free on the non-degenerate graph maps out of the discrete $n$-cube $Q^n$, so the theory is singular in the same sense as the singular homology of spaces.
It was defined first for finite metric spaces \cite{barcelo-capraro-white} and then for graphs, where the first two homology groups of several families were computed and vanishing results obtained in higher degrees \cite{barcelo-greene-jarrah-welker:comparison,barcelo-greene-jarrah-welker:vanishing,barcelo-greene-jarrah-welker:connections}.
It forms one piece of a wider combinatorial program, discrete homotopy theory, which grew out of an analysis of connectivity in social structures \cite{atkin:i,atkin:ii,kramer-laubenbacher} and was given its present form in \cite{babson-barcelo-longueville-laubenbacher,barcelo-laubenbacher}.
That program has since been recast in homotopy-theoretic terms \cite{carranza-kapulkin:cubical-graphs}, and the discrete and classical homotopy theories are now known to agree after localizing at $n$-equivalences, for every $n$ \cite{carranza-kapulkin:n-types}.

The invariant is not merely formal, and it has been put to work in several directions.
Basis graphs of matroids have trivial discrete fundamental group \cite{maurer:basis-graphs}, and since the first discrete homology group is the abelianization of that group \cite[Theorem~4.1]{barcelo-capraro-white}, a non-vanishing $H_1$ certifies that a graph is not the basis graph of any matroid, which is a condition one can check by computation.
In the theory of subspace arrangements, the fundamental group of the complement of a $k$-equal arrangement is a discrete fundamental group of an order complex, so a discrete invariant of an explicitly presented graph computes a topological invariant of an arrangement \cite{barcelo-severs-white}.
Further connections, to $k$-connectivity of graphs and to $q$-analysis of simplicial complexes, are surveyed in \cite{barcelo-laubenbacher}, and the same theory is studied in digital topology, where the objects are digital images rather than abstract graphs \cite{jamil-ali:digital}.
Most recently the invariant has found a use in topological data analysis \cite{kapulkin-kershaw:data-analysis}, and it is that application which concerns us here.

\begin{figure}[htbp]
  \centering
  \begin{tikzpicture}[
      vtx/.style={circle, fill=black, inner sep=1.6pt},
      lbl/.style={font=\small}
    ]
    % C_3
    \begin{scope}[xshift=0cm]
      \foreach \i in {0,1,2} {\coordinate (a\i) at ({90+120*\i}:1.05);}
      \fill[black!12] (a0) -- (a1) -- (a2) -- cycle;
      \draw[thick] (a0) -- (a1) -- (a2) -- cycle;
      \foreach \i in {0,1,2} {\node[vtx] at (a\i) {};}
      \node[lbl] at (0,-1.75) {$C_3$};
    \end{scope}
    % C_4
    \begin{scope}[xshift=3.6cm]
      \foreach \i in {0,1,2,3} {\coordinate (b\i) at ({90+90*\i}:1.05);}
      \fill[dhcol!25] (b0) -- (b1) -- (b2) -- (b3) -- cycle;
      \draw[thick] (b0) -- (b1) -- (b2) -- (b3) -- cycle;
      \foreach \i in {0,1,2,3} {\node[vtx] at (b\i) {};}
      \node[lbl] at (0,-1.75) {$C_4$};
    \end{scope}
    % C_5
    \begin{scope}[xshift=7.2cm]
      \foreach \i in {0,1,2,3,4} {\coordinate (c\i) at ({90+72*\i}:1.05);}
      \draw[thick] (c0) -- (c1) -- (c2) -- (c3) -- (c4) -- cycle;
      \foreach \i in {0,1,2,3,4} {\node[vtx] at (c\i) {};}
      \node[lbl] at (0,-1.75) {$C_5$};
    \end{scope}
  \end{tikzpicture}
  \caption{The shortest cycle that survives in $H_1$.
    Grey marks a cycle filled by both theories, blue one filled by discrete homology alone, and no fill a cycle filled by neither.
    A $4$-cycle bounds discretely but not simplicially, and only from length $5$ onward do the two theories agree that there is a hole.}
  \label{fig:cycles}
\end{figure}
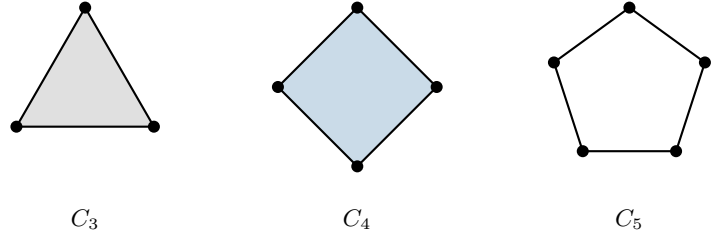

The comparison that matters there is with the Vietoris--Rips complex, whose persistent homology \cite{edelsbrunner-letscher-zomorodian:persistence,zomorodian-carlsson:computing-persistent-homology} is the standard tool of the subject \cite{carlsson:topology-and-data,otter-et-al:roadmap}.
What sets discrete homology apart from the homology of that complex is the length of the shortest cycle it still regards as a hole.
The clique complex of a graph fills in every triangle, so a $3$-cycle bounds while a $4$-cycle does not.
Discrete homology fills in the triangles and the squares alike, and the shortest cycle it leaves open therefore has length $5$; see \cref{fig:cycles}.
This one-step shift is worth a great deal on noisy data.
Four points whose pairwise distances are only approximately known will routinely produce a square whose diagonals are slightly longer than its sides, and every such square becomes a spurious bar in the Vietoris--Rips barcode.
The effect is at its worst when the input is not a metric at all, as happens for correlations between time series or for any similarity score that violates the triangle inequality, since in its absence, nothing constrains how long these bars may last.
Discrete homology never sees them.
The effect has been measured: on a noisy sample of a circle the discrete barcode carries four short bars against seventeen for the clique complex, and across a thousand randomized perturbations of a circle the discrete persistence diagram was the closer of the two to that of the unperturbed circle in $943$ trials \cite{kapulkin-kershaw:data-analysis}.

Interest in the theory has grown on both fronts.
On the computational side, the second homology group was brought within reach of machine computation by generating cubes inductively, quotienting the chain complex by the hyperoctahedral group action of \cite{greene-welker-wille}, and discarding dominated vertices before computing anything \cite{kapulkin-kershaw}, and dedicated treatments of the first \cite{ender-kapulkin:first} and of higher \cite{ender-kapulkin:higher} homology groups have followed.
On the theoretical side, filtering the discrete cubical chain complex by degree yields a spectral sequence relating discrete homology to a variant built from injective cubes, and this computes the second homology group of the Greene spheres $G^{sph}_n$ \cite{jamil-behrens:quasi-monophobic}.
What has not moved is the fourth homology group of graphs of any real size, and many third homology groups of larger or more dense graphs.
The reason is arithmetic rather than algorithmic.
The Greene sphere has ten vertices and admits roughly $2 \times 10^7$ singular $4$-cubes and more than $6 \times 10^{12}$ singular $5$-cubes, and the latter are exactly what a computation of $H_4$ needs.
Every method that begins by writing down a boundary matrix has lost before it starts.

The neighbouring homology theories of graphs have been attacked directly, and the contrast is instructive.
Path homology of digraphs \cite{grigoryans} has a persistent version with an algorithm \cite{chowdhury-memoli:path}, improved substantially in degree $1$ by identifying what a class in that degree looks like \cite{dey-li-wang:path}.
Those gains come from understanding a fixed low degree well enough to specialize, which is what makes a dedicated algorithm possible.
Ours come from attacking the enumeration itself, which is what the singular construction forces and what lets us move in the degree rather than within it.

Our contribution is an algorithm that avoids writing that matrix down.
It rests on the \redzed{} algorithm of \cite{kershaw-kapulkin:redzed}, which replaces a filtration of chain complexes by a quasi-isomorphic sequence of complexes with zero differentials and maintains the comparison maps incrementally.
The point of that reformulation is a technique it makes available, \emph{active enumeration}: a generator whose faces have all already died creates a homology class one degree above the range of interest, so it can be skipped, and the reformulation is precisely what lets one recognize such a generator before constructing it.
In the Vietoris--Rips setting, this skips over $99.7\%$ of the top-dimensional birth simplices.
For discrete homology the proportion is larger still, because the top-dimensional cubes outnumber everything below them by a wider margin, and the payoff is correspondingly greater.
The transfer is not immediate.
An $(n+1)$-simplex is obtained from an $n$-simplex by adding a vertex, whereas an $(n+1)$-cube is obtained by pairing two $n$-cubes as opposite faces, so the enumeration has to be reorganized around pairing rather than extension.
We do this in \cref{sec:red_dh} for ordinary discrete homology, giving the \redzeddh{} algorithm, and in \cref{sec:persistent} for the persistent theory, giving \redzedpdh{}.

The gains are large enough to change what is computable.
Against the algorithm of \cite{kapulkin-kershaw}, \redzeddh{} improves the computation of $H_3(T^3)$ by a factor of roughly $1700$ and that of $H_4(C_5)$ by a factor of over $250\,000$.
More to the point, it settles groups that were previously out of reach.
We obtain $H_4(G^{sph}) = 0$ in $129$ seconds, against an estimated $2 \times 10^8$ seconds for the previous method; the underlying cubical set has some $6.4$ trillion cubes, which is over $1.6$ petabytes if one were to store it, so this computation is only possible because almost none of them are ever enumerated.
%The same speedup lets us attack the Suspension Problem, which asks whether the suspensions $\Sigma_{n+1} X$ and $\Sigma_n X$ of a graph become weakly homotopy equivalent once $n$ is large, and which appears as part of Conjecture 2.1 on the AIM problem list for discrete and combinatorial homotopy theory \cite{aim:combhomotop}; see \cref{conj:suspension}.
%What a homology computation can reach is its shadow, 
The same speedup lets us attack an open conjecture in discrete homology, \cref{conj:sus} below, which asserts that an arbitrary suspension of graphs shifts discrete homology in the way that suspension of spaces shifts singular homology.
The smallest open instances are third homology groups of suspensions, and these were not computable before, since suspending a graph makes it both larger and denser.
We compute $H_3(\Sigma_3 G)$ for a range of graphs, obtain the previously unknown $H_3(\Sigma_3 G^{sph})$, and find no counterexample.
%We also verify that the collapse map $\Sigma_4 C_5 \to \Sigma_3 C_5$ induces an isomorphism on $H_4$, both groups being zero, which is the first check of that map in a degree where its behaviour was not already forced.

In the persistent setting the improvement is one of speed rather than of range.
Persistent discrete homology was proposed in \cite{kapulkin-kershaw:data-analysis} as a noise-resistant alternative to Vietoris--Rips persistence, but the obstruction to it was cost, since a filtration contains far more squares than triangles.
Active enumeration removes the obstruction.
On noisy and non-metric data almost every square is a birth square and is therefore skipped, and what remains is an algorithm competitive with the simplicial one.
On uniformly random points in $[0,1]^{10}$ our runtimes stay within a small factor of the simplicial method while producing generally under $2\%$ as many bars, and on random distance matrices, the non-metric case that motivated the theory in the first place, \redzedpdh{} is \emph{faster} than the simplicial algorithm of \cite{kershaw-kapulkin:redzed}, which is itself faster than Ripser \cite{bauer:ripser} on such input.
Together with \cite{kapulkin-kershaw:data-analysis}, this makes the case that on noisy data discrete homology is not merely the more robust choice but also the cheaper one.

The paper is organized as follows.
\cref{sec:prelims} reviews discrete homology, the computational techniques of \cite{kapulkin-kershaw}, and the \redzed{} algorithm.
\cref{sec:red_dh} presents \redzeddh{} and \cref{sec:persistent} presents \redzedpdh{}.
\cref{sec:experiments} reports the experiments, and \cref{sec:conclusion} summarizes and lists open problems.

\section{Preliminaries} \label{sec:prelims}

This section reviews what we need from three sources: discrete homology itself, the computational techniques developed for it in \cite{kapulkin-kershaw}, and the \redzed{} algorithm.
We keep the review brief and refer the reader to \cite{barcelo-greene-jarrah-welker:comparison,barcelo-capraro-white} for discrete cubical homology, to \cite{kapulkin-kershaw} for the computational techniques, and to \cite{kershaw-kapulkin:redzed} for \redzed{}.

\subsection{Discrete homology}

Discrete homology is a homology theory for simple undirected graphs, which for us means the following.

\begin{definition} \label{def:graph}
  A \textit{graph} $G = (G_V, G_E)$ consists of a vertex set $G_V$ together with an edge set $G_E \subseteq G_V \times G_V$ that is symmetric and reflexive.
  A \textit{graph map} $f \colon G \to H$ is a set map $f \colon G_V \to H_V$ such that $(fv, fw) \in H_E$ whenever $(v,w) \in G_E$.
\end{definition}

We write $v \sim w$ for $(v,w) \in G_E$.
A graph map is thus a function on vertices that carries edges to edges, and the reflexivity condition matters here: since every vertex carries a loop, a graph map is free to contract an edge by sending both of its endpoints to the same vertex.

The graphs that generate the theory are the discrete cubes.

\begin{example} \label{ex:n-cube}
  The \textit{discrete $n$-cube} $Q^n$ has vertex set $Q^n_V = \{(x_1,\dots,x_n) \mid x_i \in \{0,1\}\}$, with $\vec{x} \sim \vec{y}$ whenever $\vec{x}$ and $\vec{y}$ are equal or differ in exactly one coordinate.
\end{example}

For $1 \leq i \leq n$ and $\varepsilon \in \{0,1\}$, the face map $\delta_i^\varepsilon \colon Q^{n-1} \to Q^n$ sends $(x_1,\dots,x_{n-1})$ to $(x_1,\dots,x_{i-1},\varepsilon,x_i,\dots,x_{n-1})$.
A \textit{singular $n$-cube} in a graph $G$, or simply an $n$-cube, is a graph map $A \colon Q^n \to G$; it is \textit{degenerate} if $A \circ \delta_i^0 = A \circ \delta_i^1$ for some $i$, and \textit{non-degenerate} otherwise.

Fix a ring $R$ and let $C_n(G)$ be the free $R$-module on the non-degenerate $n$-cubes in $G$.
The map $\partial_n \colon C_n(G) \to C_{n-1}(G)$ defined on generators by
\[ \partial_n (A) = \sum_{i=1}^n (-1)^i \left[ A \circ \delta_i^0 - A \circ \delta_i^1 \right] \]
squares to zero \cite{barcelo-capraro-white}, so $(C_\bullet, \partial_\bullet)$ is a chain complex.
Its $n$-th homology group $H_n(G) = \ker \partial_n / \image \partial_{n+1}$ is the \textit{$n$-th discrete homology group} of $G$.
Equivalently, it is the homology of the cubical set $N_1(G)$ whose $n$-cubes are the singular $n$-cubes in $G$.

We now recall two of the techniques of \cite{kapulkin-kershaw}.
Our implementation uses almost all of them, as described in \cref{sec:red_dh}, but these two are the ones that shape the algorithm.

\paragraph{Pairing $n$-cubes.}
Finding the $n$-cubes is the bottleneck in any naive approach.
Done directly, one enumerates all set maps $Q^n_V \to G_V$ and keeps those that carry edges to edges, which is $|G_V|^{2^n}$ checks since $|Q^n_V| = 2^n$.
On anything short of a complete graph, essentially all of these checks fail, so the work is almost entirely wasted.

Pairing builds the $n$-cubes out of the $(n-1)$-cubes instead, using the fact that $Q^n$ is two copies of $Q^{n-1}$ with matching vertices joined.
Writing $L_n$ for the set of $(n-1)$-cubes, a pair $(A,B) \in L_n \times L_n$ determines an $n$-cube $A * B$ exactly when $A(v) \sim B(v)$ for every $v \in Q^{n-1}_V$.
Starting from $L_0 = G_V$ and iterating, one generates the cubes degree by degree, and in general at a small fraction of the cost.

\paragraph{Quotienting by the hyperoctahedral group.}
The second technique, based on \cite{greene-welker-wille}, exploits the automorphism group $\textsf{Aut}(Q^n) \cong S_n \ltimes (\mathbb{Z}_2)^n$.
An element $s \in S_n \ltimes (\mathbb{Z}_2)^n$ is a pair $s = (\sigma, r)$ with $\sigma \in S_n$ a permutation of the coordinates and $r$ a tuple of signs $r_i \in \{-1,1\}$ recording which coordinates are reversed.
Its \textit{sign} is $\textsf{sgn}(s) = \textsf{sgn}(\sigma) \cdot \prod_i r_i$, where $\textsf{sgn}(\sigma)$ is the determinant of the associated permutation matrix.
Inside the chain complex $C_\bullet(G)$ sits the subcomplex $R_\bullet(G)$ generated in degree $n$ by the elements
\[ A - \textsf{sgn}(s)\, A \circ s, \qquad A \in C_n(G), \ s \in S_n \ltimes (\mathbb{Z}_2)^n. \]

\begin{theorem}[\cite{greene-welker-wille}] \label{th:quotient}
  Let $G$ be a graph, let $n \in \mathbb{N}$, and let $p$ be a prime not dividing $(n+1)!$.
  Then $H_n(G;\mathbb{F}_p) \iso H_n(C_\bullet(G)/R_\bullet(G);\mathbb{F}_p)$. \qed
\end{theorem}

The cost of this is that computing $H_n$ forces us to work over $\mathbb{F}_p$ with $p > n+1$.
The return is that the boundary matrices are indexed by orbits of cubes rather than by cubes, which cuts their size by a factor of up to $n! \cdot 2^n$.
There is a second return as well, in that the quotient lets us discard a wider class of cubes than the degenerate ones.

\begin{definition} \label{def:semidegen}
  An $n$-cube $A$ in a graph $G$ is \textit{semi-degenerate} if $A = A \circ s$ for some $s \in S_n \ltimes (\mathbb{Z}_2)^n$ of negative sign.
\end{definition}

A semi-degenerate cube satisfies $A = -A$ in the quotient, and since the characteristic is odd this forces $A = 0$.
Every degenerate cube is semi-degenerate, so discarding the latter subsumes the former and removes more.

\subsection{\redzed{}}

The starting point of \redzed{} is a change of perspective on persistent homology \cite{edelsbrunner-letscher-zomorodian:persistence,zomorodian-carlsson:computing-persistent-homology}.
Rather than reducing the boundary matrices of a filtration, one replaces the filtration by a quasi-isomorphic sequence of chain complexes whose differentials vanish, so that the homology can simply be read off.
The filtrations to which this applies are the finest ones.

\begin{definition}
  Let $F$ be a field.
  An \textit{elementwise} filtration of chain complexes over $F$ is a sequence
  \[ C^0 \hookrightarrow C^1 \hookrightarrow C^2 \hookrightarrow \dots \hookrightarrow C^k \]
  such that for each $1 \leq i \leq k$ there is a $p$ with $(C^i/C^{i-1})_n = F$ for $n = p$ and $0$ otherwise.
  In other words, each inclusion adds exactly one generator, in exactly one degree.
\end{definition}

\begin{theorem}[{\cite[Theorem~2.1]{kershaw-kapulkin:redzed}}] \label{th:reduction}
  Let $F$ be a field and let $C^* \colon 0 = C^0 \hookrightarrow C^1 \hookrightarrow \dots \hookrightarrow C^m$ be an elementwise filtration of chain complexes over $F$.
  Then there is a diagram
  \begin{center}
  \begin{tikzcd}
  C^0 \arrow[r] \arrow[d, "r^0"' , "\simeq"]
    & C^1 \arrow[r] \arrow[d, "r^1"', "\simeq"]
    & \cdots \arrow[r]
    & C^m \arrow[d, "r^m"', "\simeq"] \\
       A^0 \arrow[r]
    & A^1 \arrow[r]
    & \cdots \arrow[r]
    & A^m
  \end{tikzcd}
  \end{center}
  in which every $r^i$ is a quasi-isomorphism and every differential $\partial^i_j \colon A^i_j \to A^i_{j-1}$ is zero. \qed
\end{theorem}

We do not reproduce the \redzed{} algorithm in full, but the shape of it follows from the theorem.
Since $A^i$ has zero differentials, $A^i \iso H_*(C^i)$, and computing $A^i$ from $A^{i-1}$ requires only $r^{i-1}$, $A^{i-1}$, $C^{i-1}$, and the generator added at time $i$.
Persistent homology is therefore computed by updating $r^i$ once per generator.
Concretely, one maintains for each degree $0 \leq p \leq n$ a pair of dictionaries $R_p$ and $R_p^{-1}$, the first a sparse representation of $r_p^i$ and the second a partial inverse kept only to speed up lookups.
Viewed from a distance, this generalizes part of the simplicial algorithm of \cite{Dey-Fan-Wang}, and it can equally be read as a matrix algorithm combining exhaustive and retrospective reduction with compression \cite[Proposition~2.11]{kershaw-kapulkin:redzed}, for which see \cite{keeping-it-sparse}.

Taken by itself, this reformulation is not an improvement; on Vietoris--Rips filtrations it is considerably slower than an implementation such as Ripser \cite{bauer:ripser}.
What makes \redzed{} worthwhile is a technique the reformulation makes available, called \textit{active enumeration}.
The bottleneck in computing Vietoris--Rips persistent homology up to degree $n$ is the $(n+1)$-dimensional birth simplices, which in the language of \cref{th:reduction} are the simplices $\sigma$ with $r\partial\sigma = 0$ at the time they are added.
Such a simplex creates a class in degree $n+1$, so it contributes nothing to what we are computing, and yet these simplices are nearly all of the simplices in the filtration.
Almost all of the running time goes into reducing them.

The standard remedy is to compute cohomology instead \cite{silva-morozov-vejdemo-johansson:dualities} and to clear births as they are identified \cite{chen-kerber:twist}, which is what Ripser does.
Active enumeration is a homological alternative that avoids not only reducing these simplices but enumerating them at all.

\begin{definition}
  Let $F$ be a field and let $C^*$ be the elementwise filtration associated to a Vietoris--Rips filtration.
  A simplex $\sigma$ is \textit{active} at time $i$ if $r^i \sigma \neq 0$.
\end{definition}

Because the dictionaries $R_p$ store only the generators with nonzero image, the active simplices are exactly the keys of $R_p$, so they cost nothing to identify.
The insight is then that a simplex $\tau$ with no active faces satisfies $r\partial\tau = 0$ and is therefore a birth simplex.
Enumerating only those $(n+1)$-simplices with at least one active face consequently skips over $99.7\%$ of the $(n+1)$-dimensional birth simplices in a typical filtration.
We write \redzedvr{} for \redzed{} together with active enumeration.

\section{The \redzeddh{} algorithm} \label{sec:red_dh}

This section presents \redzeddh{}, an algorithm for computing ordinary, that is non-persistent, discrete homology of a graph.

At first sight \redzed{} is the wrong tool, since it takes a filtration as input and returns persistent homology.
The graph, however, supplies a filtration of its own: order the orbits of non-degenerate $n$-cubes however the generation procedure happens to find them, and add them one at a time.
The classes that survive to the end of this filtration, namely those whose persistence pairs are infinite, are exactly the classes of $H_n(G)$.
Discarding the finite pairs therefore turns \redzed{} into an algorithm for ordinary discrete homology.

By itself this buys little, just as it buys little for Vietoris--Rips filtrations, where the gains arrive only once active enumeration is switched on.
The situation here is the same, and if anything more extreme.
Consider the computation of $H_3(G^{sph})$.
The graph admits $21{,}857{,}554$ cubes in degrees $0$ through $4$, of which $21{,}767{,}664$ are non-degenerate and $21{,}724{,}536$ are $4$-dimensional births.
More than $99.8\%$ of the cubes processed thus contribute nothing to the answer.
Quotienting by the hyperoctahedral group changes the numbers, but the problem remains.

Active enumeration transfers to this setting, though its implementation does not.
The goal is unchanged: avoid enumerating any $(n+1)$-cube whose faces are all inactive.
What changes is how the top-dimensional generators are built.
An $(n+1)$-simplex is an $n$-simplex with a vertex adjoined, so in \redzedvr{} one extends a single active simplex.
An $(n+1)$-cube, by contrast, is a pair of $n$-cubes glued in as opposite faces, and there is no single cube to extend.

We therefore reorganize the search around pairing, exploiting the fact that we are not constrained to any specific ordering of $(n+1)$-cubes and may process them in any order.
Given an active $n$-cube $A$, we attempt to pair it with every other $n$-cube in turn, processing each $(n+1)$-cube as it is found.
The moment $A$ becomes inactive, we abandon the search and move to the next active $n$-cube.
Doing this ensures every orbit with at least one active face is found, so once every active $n$-cube has been treated this way, the computation is complete.

Before the main algorithm we need the cube generation routine \texttt{generate\_next\_cubes}.
It assumes three auxiliary functions.
The first, \texttt{hyper\_orbit}, takes an $n$-cube $A$, presented as an array of $2^n$ vertices, and returns its orbit under the hyperoctahedral group; this is computed as in \cite{kapulkin-kershaw}.
Only one representative per orbit is stored, namely the lexicographically minimal element.
The second, \texttt{rep\_and\_sign}, takes an $n$-cube and returns its orbit representative together with the sign of $A$ within its class, returning sign $0$ if $A$ is degenerate or semi-degenerate in the sense of \cref{def:semidegen}.
We use it both here and when computing signed boundaries, where each face is replaced by its representative and its coefficient multiplied by the sign.
The third, \texttt{is\_pair\_n\_cube}, decides whether a pair of $(n-1)$-cubes $A$ and $B$ pairs to an $n$-cube $A*B$.

\begin{balgorithm}
\caption{\texttt{generate\_next\_cubes}}
\begin{algorithmic}[1]
\Statex \textbf{Input:} a list $L_{low}$ of all orbit representatives of $(n-1)$-cubes
\Statex \textbf{Output 1:} a list $L$ of all orbit representatives of $n$-cubes
\Statex \textbf{Output 2:} a list $C$ of all orbit representatives of non-degenerate and non-semi-degenerate $n$-cubes
\Statex \hspace*{-\leftmargin}\hrulefill
\State Initialize empty lists $L$, $C$

\For{$i=1$ to length($L_{low}$)}
\State $A=L_{low}[i]$
\For{$j=i$ to length($L_{low}$)}
\For{$B \in \texttt{hyper\_orbit}(L_{low}[j])$}
\If{\texttt{is\_pair\_n\_cube}($A$,$B$)}
\State $Rep$, $sgn$ $= \texttt{rep\_and\_sign}(A*B)$
\State Append $Rep$ to $L$ if $Rep \notin L$
\If{$sgn \neq 0$}
\State Append $Rep$ to $C$ if $Rep \notin C$
\EndIf
\EndIf
\EndFor
\EndFor
\EndFor
\State \Return $L$, $C$
\end{algorithmic}
\end{balgorithm}

The outer loop runs over orbit representatives of $(n-1)$-cubes rather than over all $(n-1)$-cubes, which is legitimate because every orbit of $n$-cubes contains a member whose $n$-th negative face is the representative of its own orbit.
The inner loop starts at position $i$ rather than at $1$, since $A*B$ and $B*A$ are $n$-cubes together and lie in the same class.
Having fixed the representative as the negative face, we must let the positive face range over a full orbit, which is why the innermost loop runs over $\texttt{hyper\_orbit}(L_{low}[j])$.
Each $n$-cube found is replaced by its orbit representative and recorded in $L$ if it is new; in practice $L$ and $C$ are sets, so duplicates cost nothing.
Representatives that are neither degenerate nor semi-degenerate are recorded in $C$ as well.
The main algorithm feeds $C$ to \redzed{} and uses $L$ to generate the next degree.

The main algorithm assumes one further function, \texttt{process\_redzed}, which takes an $n$-cube and processes it as the next element of the filtration, updating $R_p$ and $R_p^{-1}$ accordingly.
Recall that $R_p$ stores the map $r^i_p \colon C^i_p \to A^i_p$ and $R_p^{-1}$ is a partial inverse kept for speed.
Since \cref{th:quotient} requires a prime larger than $n+1$, our version of base \redzed{} is adapted from that of \cite{kershaw-kapulkin:redzed} to run over an arbitrary $\mathbb{F}_p$ rather than over $\mathbb{F}_2$.
The one substantive change is that a sparse vector becomes a dictionary from generators to coefficients rather than a set of generators, and symmetric difference is replaced by sparse addition mod $p$.

\begin{balgorithm}
\caption{\redzeddh{}}
\begin{algorithmic}[1]
\Statex \textbf{Input 1:} a graph $G$
\Statex \textbf{Input 2:} the max dimension $n$
\Statex \textbf{Output:} a list of the Betti numbers $[\beta_0 ,\dots ,\beta_n]$
\Statex \hspace*{-\leftmargin}\hrulefill
\State Initialize empty dictionaries $R_0,\dots, R_n$ and $R^{-1}_0 , \dots ,R_n^{-1}$
\State $L=G_V$
\State $C=G_V$
\For{$A \in C$} \Comment{Zero dimensional processing}
\State $\texttt{process\_redzed}(A)$
\EndFor
\For{$p=1$ to $n$} \Comment{Normal processing up to $n$}
\State $L,C = \texttt{generate\_next\_cubes}(L)$
\For{$A \in C$}
\State $\texttt{process\_redzed}(A)$
\EndFor
\EndFor
\State $\textsf{Active} = \keys(R_n)$
\State Initialize empty list $\textsf{Processed}$
\For{$A \in \textsf{Active}$} \Comment{Top-level processing with active enumeration}
\If{$A \in \keys(R_n)$}
\For{$B \in L$}
\For{$B' \in \texttt{hyper\_orbit}(B)$}
\If{$\texttt{is\_pair\_n\_cube}(A,B')$}
\State $Rep, sgn = \texttt{rep\_and\_sign}(A*B')$
\If{$Rep \notin \textsf{Processed}$ and $sgn \neq 0$}
\State $\texttt{process\_redzed}(Rep)$
\State Append $Rep$ to $\textsf{Processed}$
\State If $A \notin \keys(R_n)$ break from loop
\EndIf
\EndIf
\EndFor
\State If $A \notin \keys(R_n)$ break from loop
\EndFor
\EndIf
\EndFor
\State \Return $[\operatorname{length}(\keys(R_p^{-1})) \text{ for } 0 \leq p \leq n]$
\end{algorithmic}
\end{balgorithm}

\redzeddh{} runs in two phases.
In the first, every cube up to degree $n$ is processed by base \redzed{}: for each $0 \leq k \leq n$ it generates the orbit representatives $L$ of $k$-cubes together with the sublist $C$ of those that are neither degenerate nor semi-degenerate, and passes each element of $C$ to \texttt{process\_redzed}.
The second phase handles degree $n+1$ by active enumeration.
It begins by recording $\textsf{Active} = \keys(R_n)$, the $n$-cubes that are active once all $n$-cubes but no $(n+1)$-cubes have been processed.
This snapshot has to be taken as a separate list, since the keys of $R_n$ change as $(n+1)$-cubes are processed and $n$-cubes die.

For each $A \in \textsf{Active}$ we first confirm that $A$ is still active by checking $A \in \keys(R_n)$; if it has since become inactive, we move on.
Otherwise we run over all $n$-cubes $B'$, obtained by taking orbits of the representatives in $L$, and test whether $A * B'$ is an $(n+1)$-cube.
When it is, we pass to its orbit representative, discard it if it is degenerate or semi-degenerate, and discard it if the orbit has already been processed.
A representative surviving all three tests is passed to \texttt{process\_redzed}.
We then re-test whether $A$ is active, and break out of both loops if it is not.

Once every active $n$-cube has been treated, the Betti numbers remain to be read off.
The generators surviving in degree $p$ are the classes of $H_p$, and $\keys(R_p^{-1})$ is a basis for that group, so $\beta_p$ is the length of $\keys(R_p^{-1})$.

\section{Persistent discrete homology with \redzed{}} \label{sec:persistent}

This section presents \redzedpdh{}, a variant of \redzedvr{} that computes the persistent discrete homology of the $1$-skeleton of a Vietoris--Rips filtration.
We specialize in two ways.
First, we compute only the first homology group, which is what makes the computation feasible.
Second, we work over $\bbF{2}$, which lets us model sparse addition by symmetric difference of sets, written $x \triangle y$.
One could instead compute over an arbitrary field with minimal changes to the algorithm.

Persistent discrete homology was proposed in \cite{kapulkin-kershaw:data-analysis} as a noise-resistant alternative to persistent simplicial homology, particularly for data that is not metric.
The reason, as \cref{thm:H1G-by-CW-complex} below makes precise, is that simplicial homology of the Vietoris--Rips complex sees a cycle as a hole once its length reaches $4$, whereas cycles are not seen by discrete homology until length $5$.
Since noise appears mainly as short cycles, killing the $4$-cycles removes a large share of it.
\cref{sec:experiments} measures how large the share is on a range of test data.

The algorithm rests on the following theorem, a consequence of \cite[Proposition~5.12]{barcelo-kramer-laubenbacher-weaver} together with the topological and discrete Hurewicz theorems \cite[Theorem~4.1]{barcelo-capraro-white}, which was used in \cite{ender-kapulkin:first} to compute non-persistent discrete homology.
For a graph $G$, let $X_G$ be the topological space obtained from $G$ (viewed as a $1$-skeletal CW complex) by attaching a $2$-cell along every simple $3$-cycle and every simple $4$-cycle.
We call the attached cells triangles and squares.

\begin{theorem}[{\cite[Theorem 3.4]{ender-kapulkin:first}}]\label{thm:H1G-by-CW-complex}
  For any graph $G$, $H_1(G) \iso H_1(X_G)$, where the former is discrete homology and the latter is singular homology. \qed
\end{theorem}

\redzedpdh{} is accordingly a modification of \redzedvr{} that fills in not only the $3$-cycles of the Vietoris--Rips graph at a given distance, as in persistent simplicial homology, but the simple $4$-cycles as well.
The filtration contains far more $4$-cycles than $3$-cycles, so at first this appears to be a much more expensive computation, and its worst-case complexity is indeed worse.
However, by using active enumeration we significantly reduce the number of $4$-cycles that ever have to be processed.
Following standard terminology, we call a cell a birth triangle or birth square if attaching it creates an $H_2$ class, and a death triangle or death square if it kills an $H_1$ class.
The distribution of death cells between triangles and squares depends on the order in which cells are processed: attaching every triangle before any square, for instance, would yield more death triangles and fewer death squares than our implementation does.

Because we compute only $H_0$, where the situation is identical to the simplicial case, and $H_1$, active enumeration here can follow \redzedvr{} more closely than it did in \cref{sec:red_dh}.
When an edge $e$ enters the filtration, we look for the $3$-cycles and $4$-cycles containing $e$ and at least one other active edge.
Viable squares with at least one other active edge are usually scarce, so we find them by iterating over the currently active edges $e'$ and checking the possible ways $e$ and $e'$ can form a cell.
The pseudocode is given below. 
We assume a function $\texttt{enumerate\_4\_cycles}(e,e')$ which returns the $4$-cycles that contain both $e$ and $e'$.

\begin{balgorithm}
\caption{\texttt{enumerate\_cells}}
\begin{algorithmic}[1]
\Statex \textbf{Input:} An edge $e = (i,j)$ to add to $G$
\Statex \textbf{Output:} A list $\textsf{checked\_cells}$ of cells which contain $e$ and another active edge to attach
\Statex \hspace*{-\leftmargin}\hrulefill
\State Initialize empty lists $\textsf{checked\_vertices}$ and $\textsf{checked\_cells}$
\For{$e' = (k,\ell) \in \textsf{Active} \setminus \{e\}$}
\If{$k \in \textsf{checked\_vertices}$ or $\ell \in \textsf{checked\_vertices}$}
\State continue
\EndIf
\If{$\{i,j\} \cap \{k,\ell\} \neq \emptyset$}
\State Mark $\textsf{common\_vertex} \in \{i,j\} \cap \{k,\ell\}$ as checked
\If{$\texttt{forms\_triangle}(e,e')$}
\State Append $\texttt{triangle}(e,e')$ to $\textsf{checked\_cells}$
\Else
\State Append each $x \in \texttt{enumerate\_4\_cycles}(e,e')$ to $\textsf{checked\_cells}$
\EndIf
\Else
\State Append $[i,j,k,l]$ or $[i,j,l,k]$ to $\textsf{checked\_cells}$ if they form a square
\EndIf
\EndFor
\State \Return $\textsf{checked\_cells}$
\end{algorithmic}
\end{balgorithm}

Triangles are searched for before squares, since they are both far fewer and cheaper to find.
If $e$ and $e'$ share a vertex then they may form a triangle; if they do, then we skip enumerating any $4$-cycles containing $e$ and $e'$ as they are not simple.
Otherwise, we attach every square containing both edges.
If $e$ and $e'$ share no vertex, only two orientations of a square are possible and both are checked directly.

Depending on how $\textsf{Active}$ happens to be ordered, our implementation may attach the same cell twice along the same boundary, for instance if a square has two or three active edges besides $e$, or attach a cell along a non-simple $4$-cycle, for instance if the edge opposing $e$ in a non-simple square is active.
Neither affects the result, since such a cell $x$ has $r\partial x = \emptyset$.
To cut down on repeated work we mark which vertices adjacent to $e$ have already been checked along an edge sharing a vertex with $e$.
If either vertex of $e'$ is already marked, the loop can immediately continue: either a triangle bisects any square that would be formed, or that square has already been enumerated.
In the implementation, $\texttt{enumerate\_cells}(e)$ runs lazily and cells are attached as they are produced, so the full list is never stored.

The main algorithm assumes two further functions.
The first, $\texttt{sort\_edges}(D)$, lists the edges in ascending order of distance, which is the order in which they enter the filtration.
The second, $\texttt{attach\_cell}(C)$, applies base \redzed{} to a cell $x$ attached along a cycle $C$, written in terms of its edges: if $r \partial x = \bigtriangle_{e \in C} R_1[e] \neq \emptyset$ then $x$ is a death cell, so we update $R_1$ and $R_1^{-1}$ and record a persistence pair in $P_1$ if the class had nonzero lifetime; otherwise, $x$ is a birth cell and does not impact $H_1$.

\begin{balgorithm}
\caption{\redzedpdh{}}
\begin{algorithmic}[1]
\Statex \textbf{Input:} A distance matrix $D$
\Statex \textbf{Output:} The persistence pairs of the filtration in dimensions $0$ and $1$
\Statex \hspace*{-\leftmargin}\hrulefill
\State Initialize empty lists $P_0$, $P_1$
\State Initialize empty dictionaries $R_0, R_1$ and $R_0^{-1},R_1^{-1}$
\State $\textsf{Active} = \keys(R_1)$
\State $\textsf{edge\_list} = \texttt{sort\_edges}(D)$
\For{$e = (i,j) \in \textsf{edge\_list}$}
\If{$R_0[i] \neq R_0[j]$} \Comment{$e$ is a death edge}
\State Update $R_0$ and $R_0^{-1}$ as in \redzed
\Else{} \Comment{$e$ is a birth edge}
\State Initialize $R_1[e] = \{e\}$ and $R_1^{-1}[e] = \{e\}$
\State \textsf{checked\_cells} = \texttt{enumerate\_cells}(e)
\For{$x \in \textsf{checked\_cells}$}
\State $\texttt{attach\_cell}(x)$
\EndFor
\If{$e \in \textsf{Active}$}
\State Search for one additional cell $x \notin \textsf{checked\_cells}$ to apply $\texttt{attach\_cell}(x)$
\EndIf
\EndIf
\EndFor
\State \Return $P_0,P_1$
\end{algorithmic}
\end{balgorithm}

The algorithm sorts the edges, then processes them one at a time as in \redzedvr{}.
After processing an edge $e$ it applies $\texttt{attach\_cell}$ to every cycle that $\texttt{enumerate\_cells}$ produces.
Afterwards, if $e$ is inactive, no further cell containing $e$ will affect the homology and we continue.
If $e$ is still active, we look for one more cell $x$ containing $e$ that has not yet been processed.
Should we find one, $e$ is necessarily the only active edge of $x$, so attaching $x$ kills $e$.
Note that the returned lists $P_0$ and $P_1$ are updated as part of $\texttt{attach\_cell}$, exactly as in base \redzed{}.

Correctness follows from \cref{thm:H1G-by-CW-complex} together with \cite[Theorem 2.6]{kershaw-kapulkin:redzed}, which establishes the correctness of \redzedvr{}.

\section{Experiments} \label{sec:experiments}

\newcommand{\oname}{\textsc{Quot\_DH}}

This section reports our experiments with \redzeddh{} and \redzedpdh{}.
Our code is written in Julia and is available at \url{https://github.com/nkershaw01/Discrete-homology-with-RedZeD}.

\subsection{\redzeddh{}}

To match the setup of \cite{kapulkin-kershaw}, the computations in this subsection were run on the Digital Research Alliance of Canada Narval cluster, on a single node with $240$ GB of memory and $64$ threads.
We begin with the test graphs of \cite{kapulkin-kershaw}, whose descriptions may be found there, and then use the improved running times to attack one of the central open conjectures in discrete homotopy theory.

\begin{figure}[htbp]
  \centering
  \begin{tikzpicture}
  \begin{groupplot}[
      group style={group size=2 by 2, horizontal sep=1.5cm, vertical sep=1.3cm,
                   xlabels at=edge bottom, xticklabels at=edge bottom,
                   ylabels at=edge left},
      width=0.44\textwidth, height=4.6cm,
      ymode=log, xmin=0.7, xmax=4.3, xtick={1,2,3,4},
      xlabel={degree $n$}, ylabel={seconds},
      label style={font=\small}, tick label style={font=\footnotesize},
      title style={font=\small}, ymajorgrids, grid style={gray!25},
      log basis y=10, every axis plot/.append style={thick},
    ]
    \nextgroupplot[title={$C_5$}, enlarge y limits={upper, value=0.3},
      legend style={font=\footnotesize, draw=none, fill=none,
        at={(0.02,0.98)}, anchor=north west, cells={anchor=west}}]
      \addplot[qline] coordinates {(1,1.71e-3) (2,4.47e-3) (3,0.637) (4,1.41e5)};
      \addlegendentry{\oname{}}
      \addplot[dhline] coordinates {(1,4.23e-5) (2,2.45e-4) (3,3.57e-3) (4,0.479)};
      \addlegendentry{\redzeddh{}}
    \nextgroupplot[title={$G^{sph}$}]
      \addplot[qline] coordinates {(1,1.91e-3) (2,1.32e-2) (3,13.8)};
      \addplot[qline,estline] coordinates {(3,13.8) (4,2e8)};
      \addplot[dhline] coordinates {(1,1.37e-4) (2,8.19e-4) (3,9.16e-3) (4,129)};
    \nextgroupplot[title={$T^3$}]
      \addplot[qline] coordinates {(1,1.98e-2) (2,1.83) (3,8.14e3)};
      \addplot[dhline] coordinates {(1,2.84e-2) (2,0.430) (3,4.83)};
    \nextgroupplot[title={$C_5^{star}$}]
      \addplot[qline] coordinates {(1,2.11e-3) (2,2.92e-2) (3,176)};
      \addplot[dhline] coordinates {(1,9.56e-5) (2,4.32e-3) (3,2.80e-2)};
  \end{groupplot}
  \end{tikzpicture}
  \caption{Time to compute $H_n$ with \oname{} and with \redzeddh{}.
    Note the logarithmic vertical scale.
    Solid segments are measured; the dashed segment is the estimate for $H_4(G^{sph})$ under \oname{}, which we did not run to completion.}
  \label{fig:dh-times}
\end{figure}
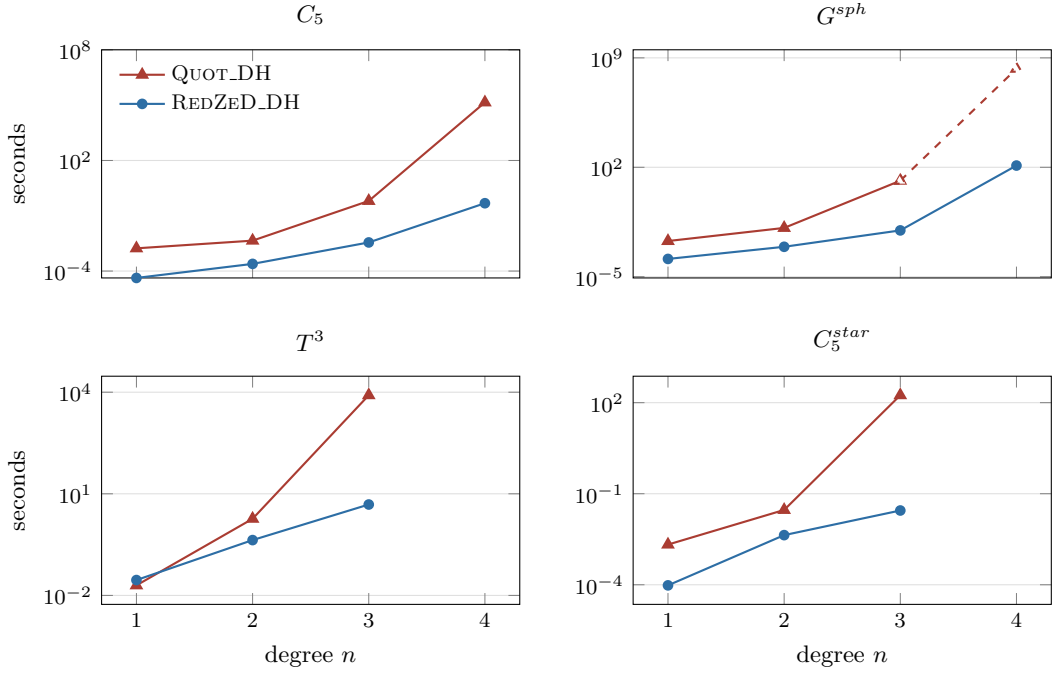

We compare against the algorithm of \cite{kapulkin-kershaw}, which we call \oname{}.
Alongside the running times we report the number of singular cubes a standard algorithm would have to enumerate, with no quotient and no active enumeration, namely the number of graph maps $Q^k \to G$ for $0 \leq k \leq n+1$.
Equivalently, this is the size of the cubical set $\operatorname{sk}_{n+1}(N_1(G))$ whose homology is being computed.
Where the exact count is itself too expensive to obtain, we give an estimate and mark it with an asterisk.
\cref{fig:dh-times} plots the running times and \cref{tab:dh-times} gives the underlying numbers.

\begin{table}[htbp]
  \centering
  \begin{threeparttable}
  \begin{tabular}{llrrrr}
    \toprule
    \textbf{Graph} & \textbf{Group} & \textbf{Cubes} & \textbf{\oname{}} & \textbf{\redzeddh{}} & \textbf{Speedup} \\
    \midrule
    $C_5$ & $H_1$ & $115$ & $1.71 \times 10^{-3}$ & $4.23 \times 10^{-5}$ & $40$ \\
          & $H_2$ & $2\,590$ & $4.47 \times 10^{-3}$ & $2.45 \times 10^{-4}$ & $18$ \\
          & $H_3$ & $990\,325$ & $0.637$ & $3.57 \times 10^{-3}$ & $178$ \\
          & $H_4$ & $8.36 \times 10^{9}$\tnote{*} & $1.41 \times 10^{5}$ & $0.479$ & $2.94 \times 10^{5}$ \\
    \midrule
    $G^{sph}$ & $H_1$ & $492$ & $1.91 \times 10^{-3}$ & $1.37 \times 10^{-4}$ & $14$ \\
              & $H_2$ & $23\,256$ & $1.32 \times 10^{-2}$ & $8.19 \times 10^{-4}$ & $16$ \\
              & $H_3$ & $21\,857\,554$ & $13.8$ & $9.16 \times 10^{-3}$ & $1.51 \times 10^{3}$ \\
              & $H_4$ & $6.44 \times 10^{12}$\tnote{*} & $2 \times 10^{8}$\tnote{*} & $129$ & $1.6 \times 10^{6}$\tnote{*} \\
    \midrule
    $T^3$ & $H_1$ & $16\,875$ & $1.98 \times 10^{-2}$ & $2.84 \times 10^{-2}$ & $0.70$ \\
          & $H_2$ & $1\,162\,250$ & $1.83$ & $0.430$ & $4.3$ \\
          & $H_3$ & $4.59 \times 10^{9}$\tnote{*} & $8.14 \times 10^{3}$ & $4.83$ & $1.69 \times 10^{3}$ \\
    \midrule
    $C_5^{star}$ & $H_1$ & $530$ & $2.11 \times 10^{-3}$ & $9.56 \times 10^{-5}$ & $22$ \\
                 & $H_2$ & $47\,570$ & $2.92 \times 10^{-2}$ & $4.32 \times 10^{-3}$ & $6.8$ \\
                 & $H_3$ & $3.08 \times 10^{8}$\tnote{*} & $176$ & $2.80 \times 10^{-2}$ & $6.29 \times 10^{3}$ \\
    \bottomrule
  \end{tabular}
  \begin{tablenotes}
    \item[*] estimated.
  \end{tablenotes}
  \end{threeparttable}
  \caption{Size of $\operatorname{sk}_{n+1}(N_1(G))$ and time in seconds to compute $H_n$.}
  \label{tab:dh-times}
\end{table}

Degrees $1$ and $2$ improve, but the interesting gains are in degrees $3$ and $4$, which is what one would expect of a technique aimed squarely at the top-dimensional generators.
For $H_3(T^3)$ the time falls from $8.14 \times 10^3$ seconds to $4.83$, a factor of roughly $1700$, and for $H_4(C_5)$ from $1.41 \times 10^5$ seconds to $0.479$, a factor of nearly $300\,000$.
The single most notable entry is $H_4(G^{sph})$, which was previously unknown.
We obtain $H_4(G^{sph}) = 0$ in $129$ seconds, against an estimated $2 \times 10^8$ seconds for \oname{}.
The size of the underlying cubical set shows why active enumeration is not merely an optimization here: it has an estimated $6.4$ trillion cubes, and storing them would take over $1.6$ petabytes.
The computation is possible only because almost none of them are ever written down.

\subsubsection{Testing the suspension conjecture}
\needspace{11\baselineskip}
\begin{wrapfigure}{r}{0.33\textwidth}
  \centering
  \vspace{-6pt}
  \resizebox{0.28\textwidth}{!}{%
  \begin{tikzpicture}[
      edge/.style={draw, line width=0.8pt},
      vertex/.style={circle, fill=black, inner sep=2pt}
  ]
  % lower layer
  \coordinate (B1) at (0,0.15);
  \coordinate (B2) at (2.4,-0.35);
  \coordinate (B3) at (1.5,-1.25);
  \coordinate (B4) at (-1.5,-1.25);
  \coordinate (B5) at (-2.4,-0.35);
  % upper layer
  \coordinate (T1) at (0,2.05);
  \coordinate (T2) at (2.4,1.55);
  \coordinate (T3) at (1.5,0.65);
  \coordinate (T4) at (-1.5,0.65);
  \coordinate (T5) at (-2.4,1.55);
  % poles
  \coordinate (N) at (0,3.15);
  \coordinate (S) at (0,-2.35);

  \draw[edge] (B1)--(B2)--(B3)--(B4)--(B5)--cycle;
  \draw[edge] (T1)--(T2)--(T3)--(T4)--(T5)--cycle;
  \foreach \i in {1,...,5}{
      \draw[edge] (B\i)--(T\i);
      \draw[edge] (S)--(B\i);
      \draw[edge] (N)--(T\i);
  }
  \foreach \i in {1,...,5}{
      \node[vertex] at (B\i) {};
      \node[vertex] at (T\i) {};
  }
  \node[vertex] at (N) {};
  \node[vertex] at (S) {};
  \node[right=4pt] at (N) {$N$};
  \node[right=4pt] at (S) {$S$};
  \end{tikzpicture}%
  }
  \caption{The suspension $\Sigma_3 C_5$.}
  \label{fig:suspension-c5}
  \vspace{-6pt}
\end{wrapfigure}
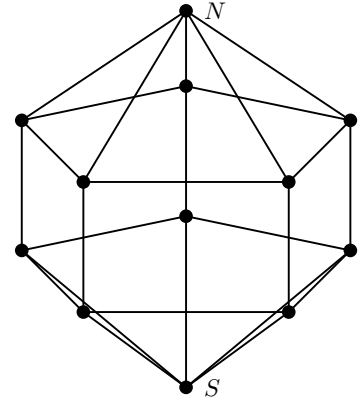
Discrete homotopy theory has no shortage of open conjectures, and one dividend of faster computation is the ability to test them.
A prime example is the Suspension Isomorphims in homology, which is only known when the length of suspension is allowed to depend on the homological degree.
%A prime example is the Suspension Problem, which asks whether suspending a graph one more time eventually stops changing its homotopy type.
%The smallest instances still open are the third homology groups of a suspension, and until now these were out of reach, since suspending a graph makes it both larger and denser.
We begin by making these statements precise.

\begin{definition}
  The \textit{box product} $G \square H$ of graphs $G$ and $H$ has vertex set $(G\square H)_V = G_V \times H_V$, with $(v,w) \sim (v',w')$ if either $v \sim v'$ and $w = w'$, or $v = v'$ and $w \sim w'$.
\end{definition}

Two pairs are thus adjacent in $G \square H$ when they agree in one coordinate and are adjacent in the other.
Suspension is built from the box product with a line.
Write $I_n$ for the line graph on $n+1$ vertices, so $(I_n)_V = \{0,1,\dots,n\}$ and $v \sim w$ if and only if $|v-w| \leq 1$.

\begin{definition}
  Let $G$ be a graph and let $n \in \mathbb{N}$.
  The \textit{length $n$ suspension} $\Sigma_n G$ is the pushout
  \begin{center}
    \begin{tikzcd}
      G\sqcup G \arrow[r, "{[i_0,i_n]}", hook] \arrow[d] & G \square I_n \arrow[d] \\
      I_0 \sqcup I_0 \arrow[r]                           & \Sigma_n G
    \end{tikzcd}
  \end{center}
  Explicitly, $(\Sigma_n G)_V = \{(v,k) \mid v \in G_V, \ 1 \leq k \leq n-1\} \cup \{N,S\}$, with $(v,k) \sim (v',k')$ if either $v = v'$ and $|k-k'| \leq 1$, or $k = k'$ and $v \sim v'$ in $G$, and with $(v,n-1) \sim N$ and $(v,1) \sim S$ for every $v$.
\end{definition}

Here $N$ and $S$ are the north and south poles.
So $\Sigma_n G$ consists of $n-1$ layers of $G$ stacked in a line, with consecutive layers joined vertex by vertex, every vertex of the top layer joined to $N$, and every vertex of the bottom layer joined to $S$.
\cref{fig:suspension-c5} shows the smallest case we use, the suspension $\Sigma_3 C_5$, which has two layers of five vertices together with the two poles, twelve vertices in all.
The picture is meant to be read as a sphere: the poles are the poles, the layers are lines of latitude, and the graph is a combinatorial stand-in for what one gets by suspending a circle.
That analogy is what the conjectures below try to make precise, and it is also why the construction is worth iterating.
%Each further suspension inserts another layer, so $\Sigma_{n+1} G$ is a longer version of the same object, and the question is whether the extra layer changes anything.
%The Suspension Problem says that eventually it does not.
%It appears as part of Conjecture 2.1 on the AIM problem list for discrete and combinatorial homotopy theory \cite{aim:combhomotop}, where it is offered as an accessible special case of the existence of homotopy colimits in the theory.

%\begin{conjecture}[Suspension Problem] \label{conj:suspension}
%  Let $X$ be a graph and let $n \geq 5$.
%  Then $\Sigma_{n+1} X$ and $\Sigma_n X$ are weakly homotopy equivalent in the sense of \cite{carranza-kapulkin:cubical-graphs}.
%\end{conjecture}

%The answer is not known for any nontrivial graph, not even for $n = 5$ and $X = C_5$.
%A weaker form, suggested by Babson, asked only that the map $\Sigma_{n+1} X \to \Sigma_n X$ collapsing a layer be $m$-connected, meaning an isomorphism on $A_k$ for all $k \leq m$, with $n$ allowed to depend on $m$, which was settled in \cite{carranza-kapulkin:n-types}, but the general statement itself remains open.

%Homology bears on this directly, since weakly equivalent graphs have isomorphic discrete homology.
%A pair $\Sigma_{n+1} X$ and $\Sigma_n X$ with $n \geq 5$ whose homology groups differ would refute \cref{conj:suspension} outright.
%Stronger, and more informative when it holds, is the statement that suspension shifts discrete homology exactly as suspension of spaces shifts singular homology.

\begin{conjecture} \label{conj:sus}
  Let $G$ be a graph and let $n \geq 3$.
  Then $H_{k+1}(\Sigma_n G) \iso H_k(G)$ for all $k \geq 1$.
\end{conjecture}

%If \cref{conj:sus} holds for $n$ and for $n+1$ then $H_{k+1}(\Sigma_{n+1} G) \iso H_k(G) \iso H_{k+1}(\Sigma_n G)$, so it implies the homological stability just described.
%A computation can therefore never confirm \cref{conj:suspension}, but it can kill it, and \cref{conj:sus} is the sharper target to aim at.

For $n > k$ the isomorphism of \cref{conj:sus} follows from the Mayer--Vietoris theorem of \cite{barcelo-capraro-white}, so the smallest counterexample one could hope for is in $H_3(\Sigma_3 G)$.
This is exactly the range that the methods of \cite{kapulkin-kershaw} cannot reach, since suspension enlarges the graph and the poles make it dense.

\redzeddh{} brings it within reach.
We test on Erd\H{o}s--R\'{e}nyi graphs on $20$ vertices with edge probability in $[0.20, 0.25]$.
The size was chosen to keep the suspensions computable while leaving room for interesting homology, and the range of $p$ was chosen empirically to maximize the chance of nontrivial $H_2(G)$.
For each graph we compute $H_2(G)$ and $H_3(\Sigma_3 G)$ and compare.
To make computations even more feasible, we use the preprocessing methods in \cite{kapulkin-kershaw} on every random graph we generate. The results can be seen below:

\begin{table}[ht]
    \centering
    \label{tab:suspension-results}
    
    \begin{tabular}{lr}
        \toprule
        \textbf{Statistic} & \textbf{Result} \\
        \midrule
        Number of graphs checked
            & 597 \\
        Graphs satisfying
            $H_3(\Sigma_3 G) \cong H_2(G)$
            & $597/597$ \\
        Average $H_3(\Sigma_3G)$ runtime per graph
            & $3255$ seconds \\
        \bottomrule
    \end{tabular}

    \vspace{0.75em}

    \begin{tabular}{cr}
        \toprule
        \textbf{Second Betti number $\boldsymbol{\beta_2(G)}$}
            & \textbf{Number of graphs} \\
        \midrule
        0 & 514 \\
        1 & 67 \\
        2 & 14 \\
        3 & 2 \\
        \bottomrule
    \end{tabular}
    \caption{Results of random graph suspension computations.}
\end{table}

We compute on 597 graphs. 
Of these 597 graphs, 514 had trivial second homology, and 83 had nontrivial second homology.
The nontrivial homology ranged from 67 with $\dim(H_2(G))=1$, 14 with $\dim(H_2(G))=2$, and 2 with $\dim(H_2(G))=3$.
All 597 graphs satisfied $H_3(\Sigma_3G) \cong H_2(G)$. 
The average computation time for $H_3(\Sigma_3G)$ was 3255 seconds, just under an hour, down from what likely would have been days to months using the previous algorithm.

We can also test for specific graphs of interest. 
The Greene sphere is the smallest known graph with nontrivial second homology, with $\dim(H_2(G^{sph}))=1$. 
The group $H_3(\Sigma_3G^{sph})$ was previously unknown, and we compute it to match $H_2(G^{sph})$ in 9 seconds.

%\fxnote{specific examples: check that $\Sigma_4 C_5 \to \Sigma_3 C_5$ is an isomorphism on $H_4$}

\subsection{\redzedpdh{}}

The computations in this subsection were run on a laptop with $16$ GB of RAM and an AMD Ryzen 7 8845H CPU.

Overall, we find that although discrete homology requires a more complicated algorithm than simplicial homology, \redzedpdh{} performs well on highly noisy data, which is where discrete homology has the greatest benefit over simplicial.
The reason is that discrete homology produces far fewer $H_1$ generators from noise, so at any given moment there are far fewer active edges to search through.
Few squares then need to be attached, and almost every birth square is avoided.
Much of the extra complexity over the simplicial algorithm therefore disappears on noisy data.
Moreover, since $\texttt{enumerate\_cells}(e)$ runs lazily as cells are attached, the squares are never stored, so memory usage is $O(n^2)$ as in \redzedvr{}.

We reuse the datasets of \cite{kershaw-kapulkin:redzed}, which we briefly recall: $\texttt{noisy\_circle}(n,\sigma)$, a unit circle sampled with $n$ points perturbed by noise drawn from a normal distribution with standard deviation $\sigma$; $\texttt{random\_euclidean}(n,d)$, consisting of $n$ points drawn uniformly from $[0,1]^d$; $\texttt{random\_distance}(n)$, a random symmetric $n \times n$ matrix with zeros on the diagonal; and $\texttt{stacked\_circles}(n,m,\sigma)$, consisting of $n$ copies of $\texttt{noisy\_circle}(m,\sigma)$ laid out in a $5 \times \frac{n}{5}$ grid.
To further test the performance on non-metric data we also use $\texttt{noisy\_matrix}(n,r)$, which directly perturbs the distance matrix of a circle with $n$ points by multiplying each off-diagonal entry by $1 + r(2X-1)$, where $X$ is drawn from a $\operatorname{Beta}(2,2)$ distribution.

\begin{figure}[htbp]
  \centering
  \begin{tikzpicture}
  \begin{axis}[
      redzedplot, width=0.62\textwidth, height=6cm,
      ymode=log, xlabel={$n$}, ylabel={seconds},
      legend columns=2,
      legend style={at={(0.5,1.03)}, anchor=south, draw=none, fill=none},
    ]
    \addplot[dhcol, mark=*, mark size=1.6pt] coordinates
      {(200,0.119226) (400,1.8167988) (600,7.5869044) (800,26.0317533)
       (1000,75.7439285) (1200,158.8914227) (1400,293.1648282) (1600,635.3503936)};
    \addlegendentry{\redzedpdh{}, $\sigma = 0.1$}
    \addplot[dhcol, dashed, mark=o, mark size=1.8pt] coordinates
      {(200,0.0417614) (400,0.4872226) (600,1.4546304) (800,7.5221118)
       (1000,13.7416538) (1200,39.3261699) (1400,29.1397472) (1600,126.7215412)};
    \addlegendentry{\redzedpdh{}, $\sigma = 0.2$}
    \addplot[vrcol, mark=square*, mark size=1.6pt] coordinates
      {(200,0.0257794) (400,0.18349) (600,0.664963) (800,2.2328795)
       (1000,4.4377188) (1200,6.6555202) (1400,10.6038333) (1600,17.1496674)};
    \addlegendentry{\redzedvr{}, $\sigma = 0.1$}
    \addplot[vrcol, dashed, mark=square, mark size=1.8pt] coordinates
      {(200,0.0269907) (400,0.1294165) (600,0.3766801) (800,0.8437295)
       (1000,1.1690647) (1200,2.4491565) (1400,2.0778207) (1600,4.0729767)};
    \addlegendentry{\redzedvr{}, $\sigma = 0.2$}
  \end{axis}
  \end{tikzpicture}
  \caption{Runtimes of $\texttt{noisy\_circle}(n,\sigma)$.
%    This is the one family in our tests where the simplicial algorithm is decisively faster, and it is also the family where the noise resistance of discrete homology is least needed.
}
  \label{fig:noisy-circle}
\end{figure}
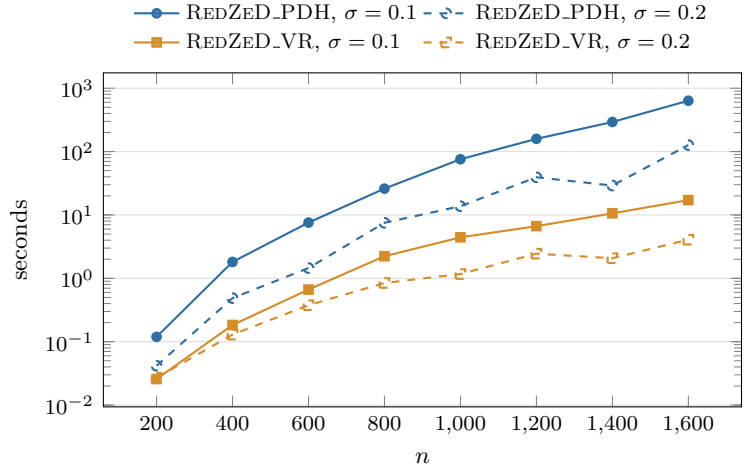

\paragraph{Noisy circles.}
We start with $\texttt{noisy\_circle}(n,\sigma)$ for $\sigma = 0.1$ and $\sigma = 0.2$; see \cref{fig:noisy-circle}.

The two choices of $\sigma$ behave quite differently, with $\sigma = 0.1$ both slower and worse scaling, for two reasons.
The first is common to every persistent method: $\texttt{noisy\_circle}(n,0.1)$ has a larger coning distance than $\texttt{noisy\_circle}(n,0.2)$, so more edges must be added before the space becomes a cone, hence contractible, and the algorithm can stop.
The second is specific to discrete homology.
The main $H_1$ class in $\texttt{noisy\_circle}(n,0.1)$ is long-lived, so many active edges all point to the same class -- around two to three times as many as for $\sigma = 0.2$.
As there are many more concurrent active edges, many squares are enumerate with each successive edge even though very few are relevant for $H_1$.
This impacts discrete methods far more than simplicial ones, since the number of squares is so much larger: up to $1{,}511{,}456{,}129$ squares were attached, an estimated $1.2\%$ of all squares in the filtration, of which only $151{,}209$ were death cells.

In terms of noise-reduction, for $\texttt{noisy\_circle}(800,0.1)$ simplicial homology produces $166$ bars against $101$ for discrete homology, and for $\texttt{noisy\_circle}(800,0.2)$ it produces $193$ against $103$.

\begin{figure}[htbp]
  \centering
  \begin{tikzpicture}
  \begin{groupplot}[
      redzedplot, width=0.42\textwidth, height=5.2cm,
      group style={group size=2 by 1, horizontal sep=1.4cm},
      ymode=log, xlabel={$n$}, xmin=100, xmax=2100,
    ]
    \nextgroupplot[ylabel={seconds}, legend columns=-1,
      legend style={at={(1.14,1.04)}, anchor=south, draw=none, fill=none}]
      \addplot[dhline] coordinates
        {(200,0.0056339) (300,0.0157705) (400,0.0279333) (500,0.0356225)
         (600,0.058896) (700,0.1353131) (800,0.1409386) (900,0.1940343)
         (1000,0.2381) (1100,0.4776445) (1200,0.3457097) (1300,0.4294278)
         (1400,0.7859483) (1500,0.7795794) (1600,0.7600201) (1700,1.3993127)
         (1800,2.0795669) (1900,1.5679531) (2000,2.0655506)};
      \addlegendentry{\redzedpdh{}}
      \addplot[vrline] coordinates
        {(200,0.0073993) (300,0.0202029) (400,0.0313504) (500,0.0322184)
         (600,0.049259) (700,0.1126202) (800,0.1035156) (900,0.1256455)
         (1000,0.1471865) (1100,0.2808515) (1200,0.1880751) (1300,0.1864001)
         (1400,0.3704305) (1500,0.3398125) (1600,0.2980936) (1700,0.5426084)
         (1800,0.7330536) (1900,0.5708163) (2000,0.6768021)};
      \addlegendentry{\redzedvr{}}
    \nextgroupplot[ylabel={bars in $H_1$}]
      \addplot[dhline] coordinates
        {(200,3) (300,4) (400,5) (500,2) (600,3) (700,9) (800,3) (900,3)
         (1000,3) (1100,10) (1200,6) (1300,3) (1400,12) (1500,8) (1600,3)
         (1700,6) (1800,14) (1900,8) (2000,12)};
      \addplot[vrline] coordinates
        {(200,78) (300,191) (400,416) (500,322) (600,281) (700,823) (800,312)
         (900,692) (1000,529) (1100,1163) (1200,550) (1300,613) (1400,1627)
         (1500,1090) (1600,439) (1700,969) (1800,2135) (1900,737) (2000,2214)};
  \end{groupplot}
  \end{tikzpicture}
  \caption{Running times, left, and number of $H_1$ bars, right, on $\texttt{random\_euclidean}(n,10)$.}
  \label{fig:random-euclidean}
\end{figure}
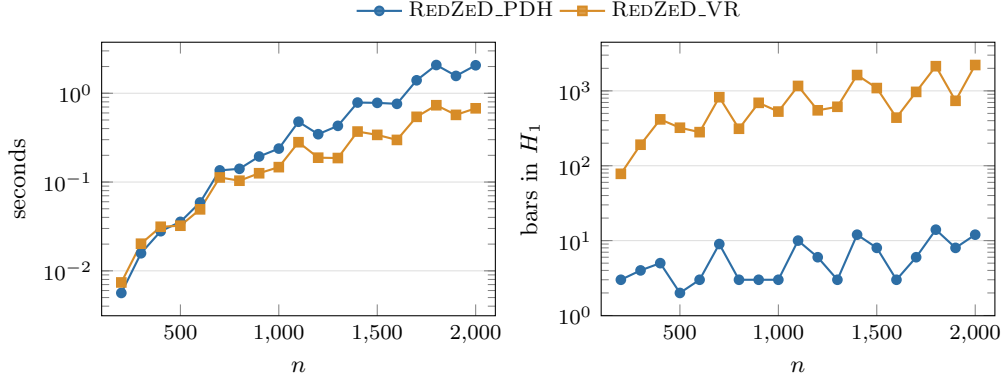

\paragraph{Random Euclidean points.}
We turn to $\texttt{random\_euclidean}(n,10)$, which is a source of pure topological noise.
Again we test both \redzedpdh{} and \redzedvr{} side by side, this time tracking both runtime and the number of persistence pairs in $H_1$; see \cref{fig:random-euclidean}.

Over this range the two running times stay within a small factor of each other, with \redzedpdh{} scaling a little worse.
The reason is that hardly any squares are enumerated at all: at $n = 2000$ only $2{,}224$ squares were attached compared to $256{,}650$ triangles, and $2{,}215$ of those squares were death squares.
The number of death squares accounts for the difference in how many persistence pairs were produced: simplicial homology produced up to $2{,}214$ bars over this range while discrete homology produced at most $14$, generally under $2\%$ as many.
This provides strong evidence for how using persistence discrete homology can reduce noise compared to simplicial.

\begin{figure}[htbp]
  \centering
  \begin{tikzpicture}
  \begin{groupplot}[
      redzedplot, width=0.42\textwidth, height=5.2cm,
      group style={group size=2 by 1, horizontal sep=1.4cm},
      ymode=log, xlabel={$n$}, xmin=25, xmax=825,
    ]
    \nextgroupplot[ylabel={seconds}, legend columns=-1,
      legend style={at={(1.14,1.04)}, anchor=south, draw=none, fill=none}]
      \addplot[dhline] coordinates
        {(50,0.000839) (100,0.0061752) (150,0.0152872) (200,0.0609083)
         (250,0.1083816) (300,0.1525561) (350,0.2422437) (400,0.3057997)
         (450,0.4424686) (500,0.5865072) (550,1.6787681) (600,2.0791396)
         (650,2.8078887) (700,3.1344352) (750,4.864202) (800,6.5035105)};
      \addlegendentry{\redzedpdh{}}
      \addplot[vrline] coordinates
        {(50,0.0013858) (100,0.0065087) (150,0.0196932) (200,0.0639718)
         (250,0.1299841) (300,0.3934457) (350,1.3534946) (400,2.2141262)
         (450,3.9641195) (500,10.1199719) (550,10.9786525) (600,24.5439064)
         (650,33.1355631) (700,53.3787387) (750,116.0300949) (800,141.7087072)};
      \addlegendentry{\redzedvr{}}
    \nextgroupplot[ylabel={bars in $H_1$}]
      \addplot[dhline] coordinates
        {(50,36) (100,115) (150,203) (200,327) (250,452) (300,616) (350,748)
         (400,892) (450,1113) (500,1286) (550,1498) (600,1637) (650,1850)
         (700,2069) (750,2285) (800,2534)};
      \addplot[vrline] coordinates
        {(50,119) (100,343) (150,659) (200,1056) (250,1478) (300,2020)
         (350,2571) (400,3071) (450,3779) (500,4403) (550,5176) (600,5869)
         (650,6819) (700,7479) (750,8269) (800,9196)};
  \end{groupplot}
  \end{tikzpicture}
  \caption{Running times, left, and number of $H_1$ bars, right, on $\texttt{random\_distance}(n)$.
    %This is the non-metric case, and the one where \redzedpdh{} overtakes the simplicial algorithm outright.
    }
  \label{fig:random-distance}
\end{figure}
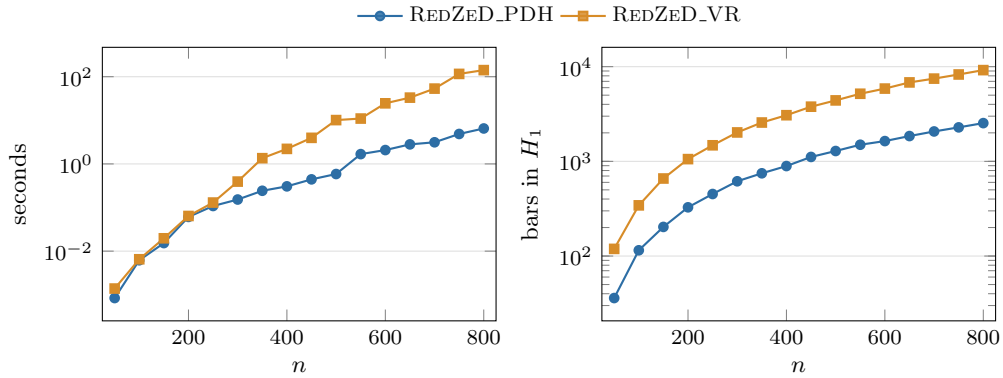

\paragraph{Random distance matrices.}
Next is $\texttt{random\_distance}(n)$, which simulates pure non-metric noise; see \cref{fig:random-distance}.

On this dataset \redzedpdh{} is actually \emph{faster} than \redzedvr{}, which was in turn shown in \cite{kershaw-kapulkin:redzed} to beat Ripser on random distance matrices.
We mention two aspects of why this occurs.
First, \redzedpdh{} produces only $27\%$ to $34\%$ as many persistence pairs.
Second, the pairs it does produce are much shorter: at $n = 800$ the average lifetime was $0.008$ for discrete homology against $0.030$ for simplicial homology, so active edges die sooner and there are fewer of them to search though at any moment.
This is in part due to the fact that without the triangle inequality, an $H_1$ class represented by a square can persist arbitrarily long using simplicial homology.
For the $n = 800$ case, \redzedpdh{} attaches $306{,}505$ triangles and $22{,}975$ squares, out of $82{,}312{,}634$ triangles and an estimated $10{,}144{,}878{,}341$ squares in the filtration, and only $14{,}100$ of the attached cells are births: $577$ from $3$-cycles and $13{,}523$ from $4$-cycles.

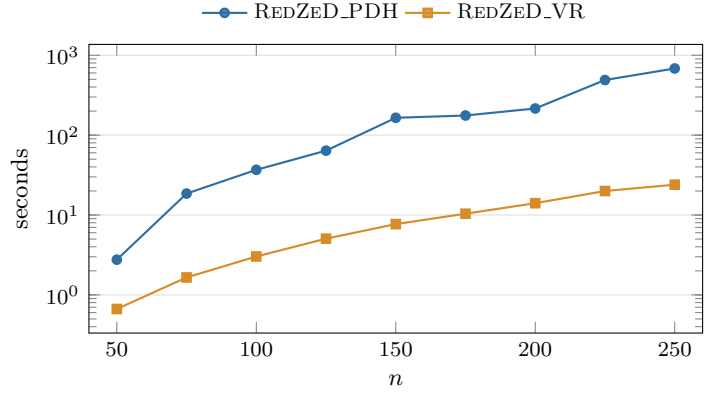
\begin{figure}[htbp]
  \centering
  \begin{tikzpicture}
  \begin{axis}[
      redzedplot, width=0.6\textwidth, height=5.4cm,
      ymode=log, xlabel={$n$}, ylabel={seconds},
      xmin=40, xmax=260, legend columns=-1,
      legend style={at={(0.5,1.18)}, anchor=north, draw=none, fill=none}
    ]
    \addplot[dhline] coordinates
      {(50,2.7592606) (75,18.5394887) (100,36.7850331) (125,63.8951763)
       (150,164.7058405) (175,176.0987987) (200,215.6053216) (225,489.6233348)
       (250,682.9133635)};
       \addlegendentry{\redzedpdh{}}
    \addplot[vrline] coordinates
       {
        (50,0.6660708)
        (75,1.6532629)
        (100,3.0261776)
        (125,5.054932)
        (150,7.6999761)
        (175,10.3848256)
        (200,14.0605715)
        (225,19.9666557)
        (250,23.9700902)
       };
       \addlegendentry{\redzedvr{}}
  \end{axis}
  \end{tikzpicture}
  \caption{Runtimes of \redzedpdh{} on $\texttt{stacked\_circles}(n,20,0.1)$.}
  \label{fig:stacked-circles}
\end{figure}

\paragraph{Stacked circles.}
The next dataset is $\texttt{stacked\_circles}(n,20,0.1)$, which simulated having many concurrent $H_1$ features; see \cref{fig:stacked-circles}.

Recall that $n$ counts the circles, which each have $20$ points, so there are up to $5000$ points in the range of $n$ considered.
While there are many active edges at the same time, as in $\texttt{noisy\_circle}(n,\sigma)$, each is short lived.
As a result, few squares are attached: in the $n = 250$ case (that is, $5000$ points), $9{,}448{,}576$ triangles and $857{,}449$ squares were enumerated.

\begin{figure}[htbp]
  \centering
  \begin{tikzpicture}
  \begin{groupplot}[
      redzedplot, width=0.42\textwidth, height=4.8cm,
      group style={group size=2 by 2, horizontal sep=1.4cm,
        vertical sep=1.3cm, xlabels at=edge bottom, xticklabels at=edge bottom},
      ymode=log, xlabel={$r$}, xmin=0.35, xmax=0.95, xtick={0.4,0.5,0.6,0.7,0.8,0.9},
    ]
    \nextgroupplot[title={$n = 100$}, ylabel={seconds},
      enlarge y limits={upper, value=0.35}, legend pos=north east]
      \addplot[dhline] coordinates
        {(0.4,0.0086847) (0.5,0.00697886) (0.6,0.00569984) (0.7,0.00445636)
         (0.8,0.00369098) (0.9,0.00351506)};
      \addlegendentry{\redzedpdh{}}
      \addplot[vrline] coordinates
        {(0.4,0.0048382) (0.5,0.00460756) (0.6,0.0045458) (0.7,0.00456998)
         (0.8,0.0045654) (0.9,0.0045824)};
      \addlegendentry{\redzedvr{}}
    \nextgroupplot[title={$n = 800$}, ylabel={seconds}]
      \addplot[dhline] coordinates
        {(0.4,21.15543958) (0.5,13.08987018) (0.6,8.39094326) (0.7,6.02999888)
         (0.8,3.01737212) (0.9,1.56637502)};
      \addplot[vrline] coordinates
        {(0.4,0.86522194) (0.5,0.7249614) (0.6,0.59912694) (0.7,0.53614854)
         (0.8,0.49808692) (0.9,0.48845718)};
    \nextgroupplot[title={$n = 100$}, ylabel={bars in $H_1$}]
      \addplot[dhline] coordinates
        {(0.4,1) (0.5,1) (0.6,1.8) (0.7,6.4) (0.8,18.2) (0.9,37.2)};
      \addplot[vrline] coordinates
        {(0.4,3.2) (0.5,16.8) (0.6,50) (0.7,93) (0.8,141) (0.9,166.2)};
    \nextgroupplot[title={$n = 800$}, ylabel={bars in $H_1$}]
      \addplot[dhline] coordinates
        {(0.4,1) (0.5,1.2) (0.6,10.2) (0.7,34.6) (0.8,103.6) (0.9,309.6)};
      \addplot[vrline] coordinates
        {(0.4,25) (0.5,102.4) (0.6,301.6) (0.7,803.8) (0.8,1720.6) (0.9,2770.8)};
  \end{groupplot}
  \end{tikzpicture}
  \caption{Running times, top, and number of $H_1$ bars, bottom, on $\texttt{noisy\_matrix}(n,r)$ at $n = 100$ and $n = 800$.}
  \label{fig:noisy-matrix}
\end{figure}
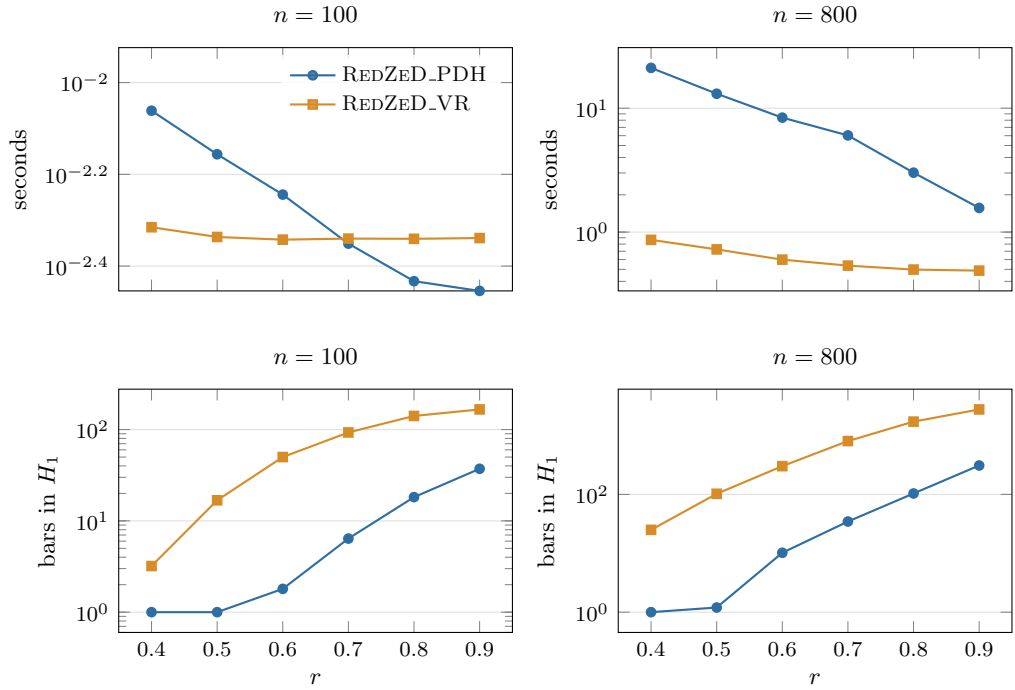

\paragraph{Noisy distance matrices.}
Finally we consider $\texttt{noisy\_matrix}(n,r)$ for varying $r$, at $n = 100$ and $n = 800$, comparing both runtime and the number of persistence pairs of an average of five trials; see \cref{fig:noisy-matrix}.

Both methods speed up as $r$ increases, which is expected with more noise, but the effect is more prominent for the discrete algorithm: at $n = 800$ its time falls by a factor of $13$ across the range while the simplicial time falls by a factor of under $2$.
The noise reduction is again significant, with discrete homology producing between $1\%$ and $31\%$ as many bars, and this reduction scales with $n$: at $n = 800$ it is at most $11\%$ and as low as $1.2\%$.

\paragraph{Summary.}
Across all of the tests, active enumeration lets us avoid attaching over $96.6\%$ of birth triangles and $97.9\%$ of birth squares, with the proportion scaling better as the number of points increases.
The worst case is a long-lived $H_1$ generator carrying a low-noise signal together with a large coning distance, as in $\texttt{noisy\_circle}(n,0.1)$.
Many edges then stay active for most of the filtration until the class representing the main cycle finally dies, which results in many birth squares having several active edges, hence being enumerated.
However, this is also a situation where persistent simplicial homology performs well, and a noise-resistant method is not needed.
On noisy and on non-metric data, discrete methods greatly reduces the noise, and \redzedpdh{} runs more comparably to the simplicial algorithm or, on pure noise, faster.
The computational difficulties encountered in $\texttt{noisy\_circle}(n,\sigma)$ are also not encountered in the other tests: for every other test, fewer than $0.1\%$ of birth triangles and $0.6\%$ of birth squares are attached, with the number of birth squares attached often below $0.01\%$.

Future improvements to \redzedpdh{} include: implementing a multi-threaded search for cells to attach, in particular for squares; calculations over arbitrary $\mathbb{F}_p$; and methods to reduce the number of birth cells enumerated in the worst case of $\texttt{noisy\_circle}(n,\sigma)$.
\section{Conclusion} \label{sec:conclusion}

We have given an algorithm for computing discrete homology, in both its ordinary and its persistent form, built on the \redzed{} philosophy of replacing a filtration by a quasi-isomorphic sequence of complexes with zero differentials.
The point of that reformulation is what it makes available downstream.
Once the comparison maps are maintained explicitly, one can tell that a generator is a birth before constructing it, and generators one degree above the range of interest can be skipped rather than reduced.
For discrete homology, where the top-dimensional cubes outnumber everything beneath them by several orders of magnitude, this is the difference between a computation and an estimate.

In the ordinary setting the payoff is new computations.
\redzeddh{} improves on the algorithm of \cite{kapulkin-kershaw} by factors reaching $10^5$ in degree $4$, and more importantly it settles groups that no previous method could reach.
Chief among them is $H_4(G^{sph}) = 0$, obtained in $129$ seconds from a cubical set with some $6.4$ trillion cubes, essentially none of which are ever enumerated.
%The same gain opens up the Suspension Problem, \cref{conj:suspension}, which appears as part of Conjecture 2.1 on the AIM problem list for discrete and combinatorial homotopy theory \cite{aim:combhomotop}.
We also provide computational evidence for the Suspension Isomorphism, \cref{conj:sus}, whose smallest open instances are third homology groups of suspensions, and suspension makes a graph both larger and denser; we compute a range of them, including the previously unknown $H_3(\Sigma_3 G^{sph})$, and find no counterexample.
%We also check that the collapse map $\Sigma_4 C_5 \to \Sigma_3 C_5$ induces an isomorphism on $H_4$, both groups vanishing, which is the first verification of that map in a degree where the answer was not already forced.

In the persistent setting the payoff is speed, in particular on the type of data discrete homology is best suited for.
Persistent discrete homology was previously shown to be more noise-resistant than simplicial, but the standing obstruction to it was cost, since a filtration contains vastly more squares than triangles.
Active enumeration removes the obstruction, because on noisy data almost every square is a birth square and is never touched.
On uniformly random points in $[0,1]^{10}$ our running times stay within a small factor of the simplicial algorithm while producing under $2\%$ as many bars, and on random distance matrices \redzedpdh{} is outright faster than \redzedvr{}, which is itself faster than Ripser \cite{bauer:ripser} on that input.
Combined with the evidence in \cite{kapulkin-kershaw:data-analysis}, this settles a question that had been open in practice: on noisy data, and especially on non-metric data, discrete homology is not only the more robust choice but also the cheaper one.

Three directions for further work suggest themselves, and they parallel those raised in \cite{kershaw-kapulkin:redzed}.

The first concerns the reach of active enumeration.
Vietoris--Rips filtrations were the first setting in which it proved decisive, and discrete homology is now the second.
Two examples make a pattern worth chasing, and the obvious candidates are the Delaunay and alpha filtrations \cite{edelsbrunner-mucke-alpha-shapes,bauer-edelsbrunner-cech-delaunay}, where the complexes are far smaller but the same asymmetry between births and deaths persists, and the multiparameter setting, where the combinatorial explosion is worse and the need correspondingly greater.
We do not know what the right general statement is, or even whether one exists.

The second is a question about what active enumeration really is.
In the end we are computing the homology of a cubical set, and skipping a generator because its faces are all inactive has the flavour of a matching in discrete Morse theory \cite{forman-morse,scoville:discrete-morse-theory}, adapted to filtrations as in \cite{mischaikow-nanda-morse}.
The analogy is suggestive but we have not made it precise.
Algebraic Morse theory \cite{kozlov:algebraic-morse} is the closest fit, since it applies to any based chain complex rather than to a cell structure, but the difficulty is the same either way: a Morse matching is a structure imposed on a complex that already exists, whereas the whole point of active enumeration is to decide what to build before building it.
Whether a matching can be produced alongside the generation rather than after it seems to us the interesting question, and an answer either way would say something about both subjects.

The third is a matter of degree.
Our persistent algorithm computes $H_1$ only, which is where \cref{thm:H1G-by-CW-complex} lets us replace cubes by cells attached along $3$-cycles and $4$-cycles.
Nothing comparable is available higher up, and a naive extension would have to work with cubes directly.
The recent progress on higher discrete homology groups \cite{ender-kapulkin:higher} is the natural place to look for a substitute, and combining it with active enumeration is the most direct route to a persistent algorithm beyond degree $1$.

\bibliographystyle{amsalphaurlmod}
\bibliography{all-refs}

%Uncomment the following if you would like an appendix
 \appendix
 \renewcommand{\thesection}{\Alph{section}}
 %\begin{appendices}
 %\section{Code for computing monoidal products}
   %\lstinputlisting[language=Python]{Graph Products.py}
 %\end{appendices}
% End of appendix
%\newpage
% Uncomment the following if you have a bibliography file

\end{document}